\documentclass[useAMS,usenatbib,referee]{raa}
\usepackage{amsmath,amssymb}
\usepackage{natbib}
\usepackage{graphicx}
\usepackage{times}

\begin{document}

\title{A quick estimation of luminosity function based on the luminosity-distance diagram}

   \volnopage{Vol.0 (200x) No.0, 000--000}      %%preserved for Editor. DOn't remove!
   \setcounter{page}{1}          %%starting page, preserved for Editor. DOn't remove!

\author{Yuan-Chuan Zou \inst{1}}

\institute{School of Physics, Huazhong University of Science and Technology, Wuhan 430074, China; zouyc@hust.edu.cn(YCZ)}

 \date{Received~~2017 month day; accepted~~2017~~month day}

\abstract{
Based on the luminosity-distance diagram, we propose a method to quickly estimate the luminosity function for any certain astrophysical objects. Giving the mean distance between any two objects at a given luminosity range, we can find the relation between the mean distance and the luminosity, and consequently can obtain the luminosity function. Not like the straightforward counting method, this method does not need a complete sample. The only requirement is that the object distributes uniformly in space.  We apply this method to a simulated sample, and find it can produce the luminosity function properly. This method can also be used for energy function.
}

\authorrunning{Zou}
\titlerunning{Luminosity function from the L-D diagram}

\maketitle

\section{Introduction}\label{sec:intro}
Luminosity function is a basis concept in the population and evolution of the astrophysical objects \citep{1999acfp.book.....L}. It is widely used in galaxies \citep{1976ApJ...203..297S}, stars \citep{1955ApJ...121..161S}, supernovae \citep{2011MNRAS.412.1441L}, $\gamma$-ray bursts \citep{2000ApJ...535..152K}, white dwarfs \citep{1988ApJ...332..891L}, etc.
The straightforward method to gain the luminosity function is to count the number at different luminosities. However, it suffers the so-called data biases for the certain object \citep[e.g.][]{1998ApJ...495..251D}, such as the Malmquist bias \citep{1920MeLuS..22....3M}, that the dimmer objects are more likely not  included in the sample. Without a complete unbiased sample, the luminosity function is not very reliable.

In reality, it is not easy to obtain a complete sample, either because the faint source is too far, or because  no complete survey project. Especially with the more and more data accumulation, more sub-classes of the objects have been found. For these incomplete samples, the luminosity function (if available) is still very helpful for understanding  the object.

Here we propose a method to gain the luminosity function with incomplete sample, only if the object is distributed uniformly in the space. Taking this advantage, we do not need to observe all the objects in the whole space, but neighbors are enough. Then we can infer the object in outer space (and also dimmer) has the same density. For objects with different luminosity, the density is different, and the neighbors are farther or nearer. We then summarize them into the luminosity function.

The basic method is shown in section {\ref{sec:method}}. We apply the method for simulated data in section {\ref{sec:app}}, and also show the steps for not well observed sample. We discuss the requirement of this method and how it can be applied to other conditions in section \ref{sec:discuss}.

\section{Method}\label{sec:method}
If the given objects are uniformly distributed in the space around the local universe, roughly, the distance to the nearest one indicates the average distance between two objects, and consequently indicates the number density. Considering the objects obey a certain luminosity function, they also distribute uniformly for a given luminosity range $(L, L+{\Delta} L)$. Also, the nearest one in the range $(L, L+{\Delta} L)$ indicates the average distance for this luminosity range.
More precise average distance can be fitted in the luminosity-distance diagram for the given luminosity range. As the uniform distribution, the accumulated number inside radius $R$ should obey $N(L,R) {\Delta}L= \frac{4\pi }{3} n(L) R^3 {\Delta}L $, where $ n(L)$ is the number density for a given luminosity range. As $N(L,R)$ is countable for any radius range $(R, R+{\Delta d}R)$, the $n(L)$ can be fitted with the counted numbers. With the fitted $n(L)$ and the prior set  range $(L, L+{\Delta} L)$, one can fit the relation between $n(L)$ and $L$, which is the luminosity function.

An alternative way is to fit the average distance $D(L)$ v.s. luminosity $L$ on the luminosity-distance diagram, where the average distance  $D(L) \equiv n(L)^{-1/3}$ is obtained from the fitted $n(L)$.
In the luminosity-distance diagram, suppose the fitted relation between $D(L)$ and $L$ is $f(L)$, one gets the luminosity function:
\begin{equation}
  n(L)  = f(L)^{-3}.
  \label{eq:luminosityFunction}
\end{equation}

\section{Application}\label{sec:app}
To verify the reliability, we apply this method to a simulated data, which has a prior luminosity function  used from comparison. For the simulated data, we have the following steps.

\begin{figure}
 \includegraphics[width=0.8\textwidth]{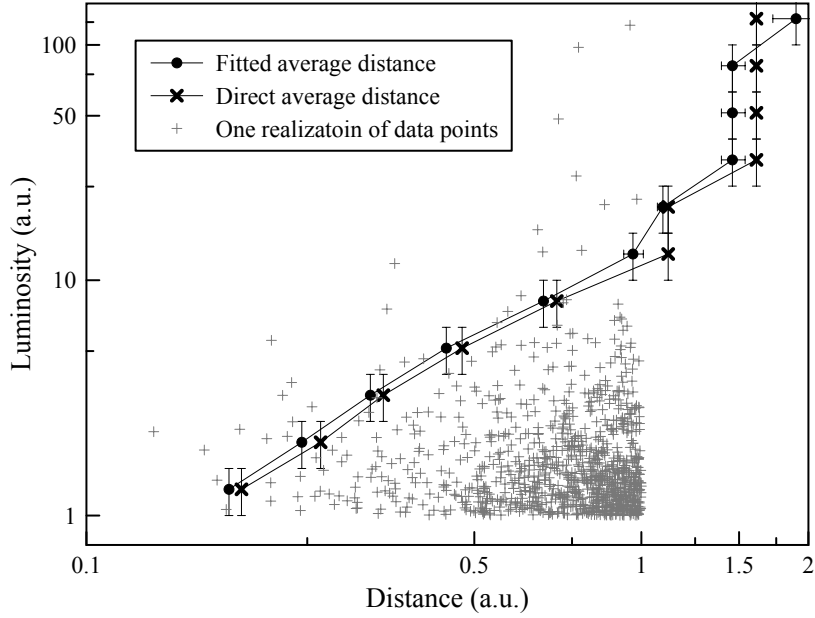}
 \caption{Luminosity-distance diagram for simulated objects with input luminosity function $n(L) \propto L^{-2}$, in arbitrary unit. The gray pluses  are the realization of total N=1000 objects, which are distributed uniformly in space inside a sphere with radius 1. Crosses  are the average distances between two objects in certain luminosity ranges, which are shown as y-axis error bars.  Black dots are the fitted average distances in the same luminosity ranges.
 %The up-right corner, there is a horizontal line for the Orange one, looks like an error bar. It is not. Because there is no data in around 10^3, the average distance becoming infinity, and the connecting line becomes a horizontal line.
 }
 \label{fig1}
\end{figure}

Step 1, sample $N=1000$ points uniformly and randomly distributed in a sphere within radius 1 (in arbitrary unit). The luminosity of each point is set randomly obeying a powerlaw distribution $n(L) {\rm d} L = C L^{-\alpha} {\rm d} L$ with $L_{\min} = 1$  (in arbitrary unit), and $C$ is the normalization factor obeying the total number  $N$, and the powerlaw index $\alpha$ is setting to be 2 here. The data of one realization is shown in figure \ref{fig1}, in grey tiny pluses.

Step 2, we bin the data under luminosity. In each luminosity bin, we can count the number of the sample, say $n_{\rm bin,L}$. The number density for this luminosity bin is then $n_{\rm bin,L}/(4 \pi/3)$, by noticing the volume is in sphere. The mean distance is then $[4\pi/(3n_{\rm bin,L})]^{1/3}$, which is plotted in figure \ref{fig1}, in crosses. Though the mean distance is directly calculated by counting the number, it should be different for different realization as the realization is randomly performed.

Step 3, we bin the data under radius for each  luminosity bin. For each radius bin, we count the number of the sample. This forms a sequence of radii and numbers at each radius bin. Remembering the sample is uniformly distributed, the accumulated number of the sample should be proportional to the radius, i.e., $N_{\rm acc} =\frac{4\pi}{3} n(L) R^3$, where $n$ is the number density. $n(L)$ can be fitted by the counted numbers. The fitted $n$ with errors is converted to the mean distance $D(L)\equiv n(L)^{-1/3}$, which is shown in figure \ref{fig1}, in black dots. One can see the fitted mean distance is well consistent with the directly derived mean distance.

Step 4, by fitting the sequence $A$ and the corresponding luminosity bin, we will get the luminosity function. Fitting to the black dots with powerlaw relation gives us \citep{2007MeScT..18.3438K}: $\log_{10} D(L)=-(0.711 \pm 0.034) + (0.66\pm 0.03) \log_{10} L$. Taking this into equation (\ref{eq:luminosityFunction}), we get the final fitted luminosity function $n(L)=10^{(2.1 \pm 0.1)} L^{-1.98 \pm 0.09}$. It is well consistent with the input.

In most real astrophysical objects, they are not as complete as the simulated data, mainly because the far objects are too dim (lower right side of the luminosity-distance diagram). Fortunately, on the upper left side of the luminosity-distance diagram, they are always brighter and less likely being omitted by the observations. For these non-complete samples, we follow the steps 1-4 with modifications:
In step 1, plot the luminosity-distance diagram according to the real data.
In step 2, we do not calculate the number density as the sample is not complete.
In step 3, we do not fit $N_{\rm acc} =\frac{4\pi}{3} n(L) R^3$ for all the radius bins as the lack of data in larger distances. Instead, we just choose  the nearer bins.
Finally, we can also obtain the luminosity function in step 4.

\section{Discussion}\label{sec:discuss}
In this letter, we proposed a quick method to estimate the luminosity function for any astrophysical objects, especially for those not completely sampled objects. The requirements of this method are: 1. There is a survey for this kind of object. 2. The objects are  uniformly distributed in the space from our local universe. 3. The luminosity and the distance are measured. We applied the method to a simulated data, and the result showed the method works well. As shown in section \ref{sec:method}, to get the luminosity function, one has two choices, either to fit $n(L)-L$ relation directly from step 3, or to fit $D(L)-L$ relation and then translating it to the luminosity function. The former is more directly, while the latter can be more intuitively shown on the diagram.

%For those not uniformly observed objects, this method can also be applied by disregarding the data outside a certain region, i.e., the maximum sphere region. Therefore, one needs the minimum sphere, of which it contains at least one object for all the luminosity range.

This method can be used for a rough estimation of the luminosity function directly based on the luminosity-distance diagram. One can draw a rough border in the diagram, which satisfies that there are nearly no objects on the left side of the border. As the left side is in nearer distance, the objects are less likely to be omitted. That means the observed nearest object is very likely the real nearest one. Considering the uniform distribution, this border at any luminosity indicates the nearest distance of the objects for the given luminosity, and it can be roughly taken as the mean distance $D(L)$. A rough luminosity function is then obtained. Therefore, for those extremely lack of data, one can still get a rough luminosity function based on equation (\ref{eq:luminosityFunction}).

It can also be used to derive the energy function for the explosive sources like supernovae, and gamma-ray bursts. For the cosmological objects, one should take the evolution into account.
One can also use the luminosity-distance diagram to justify whether the objects in the diagram are from the same class. In general, the luminosity function for the same class of object should obey some simple function, like log-normal, powerlaw, cutoff powerlaw etc. At least, they are all smooth. Therefore, from the luminosity-distance diagram, if the left border region is not smooth, one may conclude that they are from different classes.

This work is supported by the National Basic Research Program of China (973 Program, Grant No. 2014CB845800), and the Chinese-Israeli Joint Research Project (Grant No. 11361140349).

\end{document}